\documentclass[apl,superscriptaddress,reprint,amssymb,aps,floatfix,showkeys]{revtex4-1}
\usepackage{graphicx}
\usepackage{amsmath}
\usepackage{dcolumn}
\usepackage{bm}
\usepackage{bbold}
\usepackage{color}
\usepackage{ulem}
\usepackage{nicefrac}
\usepackage{tabularx}
\usepackage{soul}

\bibstyle{natbib}

\begin{document}

\title{Measurements of Elastic Properties of Langatate at Liquid Helium Temperatures {for design of ultra low loss mechanical systems}}

\author{Maxim Goryachev}
\email{maxim.goryachev@uwa.edu.au}
\affiliation{ARC Centre of Excellence for Engineered Quantum Systems, University of Western Australia, $35$ Stirling Highway, Crawley WA, 6009, Australia}

\author{Philippe Abb\'{e}}
\affiliation{Department of Time and Frequency, FEMTO-ST Institute, ENSMM, $26$ Chemin de l'\'{E}pitaphe, $25000$, Besan\c{c}on, France}

\author{Bernard Dulmet}
\affiliation{Department of Time and Frequency, FEMTO-ST Institute, ENSMM, $26$ Chemin de l'\'{E}pitaphe, $25000$, Besan\c{c}on, France}

\author{Roger Bourquin}
\affiliation{Department of Time and Frequency, FEMTO-ST Institute, ENSMM, $26$ Chemin de l'\'{E}pitaphe, $25000$, Besan\c{c}on, France}

\author{Serge Galliou}
\affiliation{Department of Time and Frequency, FEMTO-ST Institute, ENSMM, $26$ Chemin de l'\'{E}pitaphe, $25000$, Besan\c{c}on, France}

\date{\today}


\begin{abstract}

We present full characterisation of acoustic wave devices based on the fully synthetic crystalline material at the liquid helium temperature range { required for the design of ultra low loss mechanical systems in many areas of research including frequency control and fundamental measurements}. Temperature coefficients of the effective elastic tensor of Langatate (LGT) in Lagrangian representation are determined for the temperature range $3.8-15$K. The Lagrangian formalism is mandatory in the analysed situation since the expansion coefficients of the LGT are still unknown at these temperatures. The measurement method involves a set of high-quality resonators of various cut angles, and uses measurements of frequency-temperature relations to extract the temperature coefficients of the elastic tensor. In addition, power sensitivity of LGT resonators at cryogenic temperatures is determined and dominant loss mechanism is identified.

\end{abstract}

\maketitle


Recently low temperature applications of piezoelectric Bulk Acoustic Wave (BAW) devices has drawn serious attention due to their extraordinary quality factors\cite{Goryachev1,ScRep,quartzPRL}. Such applications include but not limited to frequency control systems\cite{PtogFreqContr}, quantum hybrid devices\cite{ScRep,Kippen,OConnell:2010fk} and atomic force spectroscopy. It is demonstrated\cite{mybook,SunFr} that in most of these applications, and in frequency control in particular\cite{perspect}, thermal response of such devices is critical and requires further improvement. For this purpose, thermal coefficient of the utilised material have to be identified at the liquid helium temperature range.

Although langasite-type materials were initially developed as laser crystals, some of their properties make them very good candidates for frequency control applications. Like other compounds of this crystal family, Lanthanum gallium tantalite (La$_{3}$Ga$_{5.5}$Ta$_{0.5}$O$_{14}$) or Langatate (LGT) belongs to the same trigonal crystal class 32 as quartz, and as a consequence exhibits similar properties.
First, it is piezoelectric with an electromechanical coupling coefficient two to three times greater than that of quartz. 
These properties make Langatate ideal for variety of frequency-control applications including BAW and surface acoustic wave (SAW) resonators and filters~\cite{Kosinski2000}. BAW LGT resonators have already demonstrated their efficiency as master resonators in very stable oscillators~\cite{ImbaudTransUFFC2008}. Since LGT-based devices exhibit similar low phase noise and high $Q$ regimes as quartz devices, they can also be classified as a low-loss material~\cite{JohnsonJAP2011}. {Nevertheless, such applications require careful selection of the material in terms of crystal quality.}
 It should also be pointed out that LGT crystal can be made according to a purely synthetic process of growth whereas quartz crystal cannot be reproduce infinitely from synthetic seeds. {Indeed, the crystalline quartz quality degenerates after a few successive growths and a new start from a natural quartz seed is needed as well.}

{Like gallium orthophosphate (GaPO$_{4}$)~\cite{KremplSA1997} or Langasite (La$_{3}$Ga$_{5}$SiO$_{14}$ or LGS), LGT is still a subject of intensive research mainly because of its ability to be used for high temperature applications. Unlike quartz, LGT does not undergo any phase transition up to the melting temperature of $1450^{\circ}$C. This makes it relevant for sensor applications in high temperature environment, e.g. inside engines.} 

A resonance frequency of any acoustic device is a function of elastic coefficients of corresponding material. Since all these coefficients are temperature dependent, the knowledge of this dependence is essential for predicting resonance frequency sensitivity to temperature and corresponding stability. Various methods have been used to determine such coefficients from phase velocity measurement by a pulse echo overlap method, or resonant ultrasound spectroscopy and combined resonance techniques~\cite{SturtevantUFFC2009, SchreuerUFFC2002}, or by means of laser ultrasonic techniques~\cite{ChillaJAP2001,Reverdy2001}. Although application of these methods at low temperatures is associated with technical problems. Alternatively, elastic coefficients can be accurately identified from the frequency-temperature measurements of plate resonators based on corresponding material~\cite{Bechman1962,Lee1983}. This technique has been used to determine LGT coefficients at room temperatures~\cite{Bourquin2006}. An extension of this last method for trapped-energy resonators operating at liquid-helium temperature is carried out in the present work.
\\


At low temperatures, the use of the Lagrande description is mandatory because the linear thermal expansion coefficients of LGT are still unknown. 
 In the Lagrange description~\cite{Thurston1974}, the material density involved in the fundamental equation of the dynamics is referred to the reference temperature in the undistorted reference state, on the surface of which the boundary conditions are also applied. According to this formalism, the stress-strain relationship is given by the following tensor expression $P_{ij} = G_{ijkl}\partial_k u_l$,
where $G_{ijkl}$ is an element of the tensor of elastic coefficients, $P_{ij}$ represents the first Piola-Kirchoff stress tensor, $u_l$ is an element of a vector of displacement. The elements are given with respect to coordinates of a material point at reference temperature. Elastic constants $G_{ijkl}$ are temperature dependent and the corresponding thermal coefficients have been already determined at room temperature\cite{Bourquin2006,Bourquin2009}.

The elastic tensor $G_{ijkl}$ determines eigenfrequencies of acoustical modes of mechanical structures. Thus, thermal sensitivity measurements of these frequencies can be used to determine thermal coefficients. 
The structure used in this approach is a BAW plate resonator whose high quality factors ensure sufficient accuracy. To resolve the material anisotropy, resonators with various cut angles should be characterised\cite{Bechman1962,Lee1983}.  

\begin{figure}[t!]
	\centering
			\includegraphics[width=0.45\textwidth]{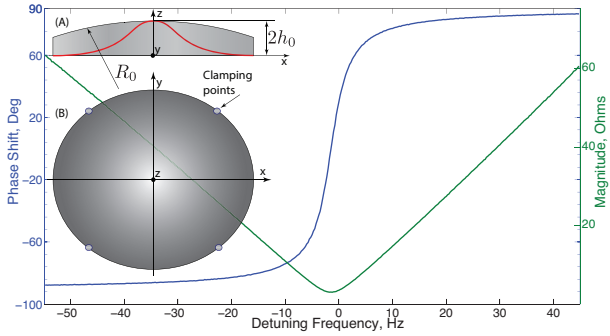}
	\caption{Frequency response of an LGT resonator in terms of its impedance $Z(f)$ is shown in the form of a phase (left vertical axis) and magnitude (right vertical axis) of measured complex values. Insets (A) and (B) are the side and top views of a plano-convex plate BAW resonator. The red curves depicts the Gaussian distribution of the acoustic energy in the plate. }
	\label{reson}
\end{figure}

The characterisation approach utilises another theoretical result, the Tiersten-Stevens theory\cite{stevens:1811} that analytically describes eigenfrequencies of contoured piezoelectric plate resonators. The theory provides the relationships between eigenfrequencies of the plate vibrations and mechanical constants that have to be determined. A case of plano-convex plate resonators of the thickness $2h_0$ and the curvature radius $R_0$, or trapped energy resonators (see insets (A) and (B) of Fig.~\ref{reson}) is considered in the study. {Predictions of the Tiersten-Stevens theory have been confirmed by many practical results collected over the last decades.}



Infinite plate acoustic thickness vibration of a certain type (shear or longitudinal) is fully determined by one wave-number that describes variation of the wave along the plate thickness, i.e. its overtone (OT), resulting in the following eigenfrequency:
\begin{equation}
	\label{R002GF}
	 f_n^2= \frac{n^2}{16h_0^2}\frac{G_e}{\rho_0}\Big(1-8\frac{k_\alpha^2}{n^2\pi^2}\Big),
\end{equation}
where $n$ is a wave-number along the $z$ axis, $\rho_0$ is the material specific mass, $2h_0$ is the plate thickness,and $G_e$ is an effective temperature dependent elastic coefficient in the direction of the wave propagation ($z$ axis in this case). $G_e$ is composed of the elements of $G_{ijkl}$ that make impact in the propagation along the chosen crystal orientation.
Both $R_0$ and $h_0$ are defined at the reference temperature that is $8$K in the present study. For high values of $n$, the piezoelectric correction, the second term in brackets, becomes negligible so that it could vanish. In the case of piezoelectric resonators excited electrically, $n$ should be odd.

{The matrix of elastic coefficients is obtained from the elastic tensor by using the compression rule of indices: $11\rightarrow1$, $22\rightarrow2$, $33\rightarrow3$, $23\rightarrow4$, $31\rightarrow5$,$12\rightarrow6$, $32\rightarrow7$, $13\rightarrow8$, $21\rightarrow9$. Thus, the elastic matrix is entirely defined by nine coefficients, in the following form: }
\begin{equation}
	\label{G006GF}
 \left( \begin{array}{llllllllllllllllll}
	G_{11} & G_{12} & G_{13} & G_{14} &0&0& G_{17}&0&0 \\
	G_{12} & G_{11} & G_{13} & -G_{14} &0&0& -G_{17}&0&0   \\
	G_{13} & G_{13} & G_{33} & 0 &0&0& 0&0&0   \\
	G_{14} & -G_{14} & 0 & G_{44} &0&0& G_{47}&0&0   \\
	0 & 0 & 0 & 0 &G_{55}&G_{17}& 0&G_{47}&G_{17}   \\
	0 & 0 & 0 & 0 &G_{17}&G_{66}& 0&G_{14}&G_{66}    \\
	G_{17} & -G_{17} & 0 & G_{47} &0&0& G_{55}&0&0   \\
	0 & 0 & 0 & 0 &G_{47}&G_{14}& 0&G_{44}&G_{14}   \\
	0 & 0 & 0 & 0 &G_{17}&G_{66}& 0&G_{14}&G_{66}    \\
 \end{array} \right),
\end{equation}
{that reflects all corresponding symmetries. In addition, $G_{66}=\frac{G_{11}-G_{12}}{2}$ for the LGT being a class $32$ element.  }

For plano-convex plate finite resonators, according to the Stevens-Tiersten analysis, an eigenfrequency of a certain vibration type is characterised by three wave-numbers $n$, $m$ and $p$:
\begin{equation}
	\label{R003GF}
	\frac{f_{nmp}^2}{f_n^2} =  1+\frac{1}{n\pi}\sqrt{\frac{2h_0}{R_0}}\Big(\sqrt{\frac{M_n}{G_e}}(2m+1)+\sqrt{\frac{P_n}{G_e}}(2p+1)\Big)
\end{equation}
where $n$, $m$ and $p$ are acoustic wave variations in $z$, $x$ and $y$ directions respectively, $f_n$ is an eigenfrequency of a corresponding infinite plate, $2h_0$ is the plate thickness, and $R_0$ is the radius of curvature. Eq.~(\ref{R003GF}) makes the measurable frequencies dependent on three temperature-dependent parameters, $G_e$, $M_n$, $P_n$.
In fact, $M_n$ and $P_n$ are much less dependent on temperature than $G_e$.

It follows from eq. (\ref{R003GF}) that the infinite plate eigenfrequency $f_n$ can be calculated from the measurement results of finite plano-convex plates $f_{nmp}$ by combining three eigenfrequencies with the same longitudinal wave-number $n$ and different combinations of in-plane wave numbers $m$ and $p$:
\begin{equation}
	\label{R004GF}
	{f_n^2} =  \frac{1}{4}\big(6f^2_{n00}-f^2_{n02}-f^2_{n20}\big).
\end{equation} 
The resultant infinite plate eigenfrequency is a function of only one temperature-dependent parameter $G_e$. Thus, it is straightforward to relate temperature dependence coefficients of the elastic parameters with that of the resonator eigenfrequencies. 

$k$th order temperature coefficients of the elastic elements and a plate eigenfrequency could be represented in the series form:
\begin{equation}
	\label{R005GF}
	\left. \begin{array}{ll}
\displaystyle \xi^{(k)}_{ij} = \frac{1}{k!}\frac{1}{G_{ij}}\partial_T^kG_{ij},
\displaystyle \chi^{(k)}_{nmp} = \frac{1}{k!}\frac{1}{f_{nmp}}\partial_T^kf_{nmp},
\end{array} \right. 
\end{equation} 
which give a coefficients of Taylor series expansions around the reference temperature $T_\text{ref}$ with the simplified (two index) notation of the elastic matrix $G_{ijkl}$\cite{Bourquin2009}. In the following only infinite plate frequency coefficients $\chi^{(k)}_{n00}=\chi^{(k)}_{n}$ are considered assuming correction (\ref{R004GF}) is imposed. This assumption is made due to the fact that $M_n$ and $P_n$ temperature dependences are negligible in comparison with that of $G_e$. The relationships between first three lowest-order coefficients for an $n$th order infinite plate eigenfrequency and effective elastic coefficient $G_e$ are given as follows:
\begin{equation}
	\label{R006GF}
	\left. \begin{array}{ll}
\displaystyle \chi^{(1)}_{n} = \frac{1}{2} \xi^{(1)}_{e},\hspace{5pt} \chi^{(2)}_{n} = -\frac{1}{8} \big(\xi^{(1)}_{e}\big)^2+\frac{1}{2} \xi^{(2)}_{e},\\
\displaystyle \chi^{(3)}_{n} = \frac{1}{16} \big(\xi^{(1)}_{e}\big)^3-\frac{1}{4} \xi^{(1)}_{e}\xi^{(2)}_{e}+\frac{1}{2} \xi^{(3)}_{e},
\end{array} \right. 
\end{equation} 
where $\xi^{(k)}_{e}$ are effective coefficients for the direction of the wave propagation. Using these algebraic relationships, it is possible to deduce material temperature coefficients using the following recurrent linearization~\cite{dulmet2001}:
\begin{equation}
	\label{R099GF}
	\left. \begin{array}{ll}
\displaystyle \chi^{(1)}_{i} = M_{ij}^{(1)} G^{(1)}_{j},
\displaystyle \chi^{(2)}_{i} = M_{ij}^{(1)} G^{(2)}_{j}+\frac{1}{2}M_{ijk}^{(2)} \xi^{(1)}_{j}\xi^{(1)}_{k},\\
\displaystyle \chi^{(3)}_{i} = M_{ij}^{(1)} G^{(3)}_{j}+M_{ijk}^{(2)} \xi^{(2)}_{j}\xi^{(1)}_{k}+\frac{1}{6}M_{ijkl}^{(3)} \xi^{(1)}_{j}\xi^{(1)}_{k}\xi^{(1)}_{l},
\end{array} \right. 
\end{equation} 
where elements $M_{ij}$ are coefficients of eigenfrequency sensitivity to elastic coefficients $M_{ij} = \frac{G_j}{f_i}\frac{\partial f_i}{\partial G_j}$.
It should be noted that $|M^{(n)}|<\frac{1}{2^n}$ and products  $\xi_j^{(1)}\xi_k^{(1)}$, $\xi_j^{(2)}\xi_k^{(1)}$ and $\xi_j^{(1)}\xi_k^{(1)}\xi_j^{(1)}$ are of the order of $10^{-12}$. Comparing this order to $10^{-6}$ for the single coefficients $\xi^{(k)}_{j}$, relationships (\ref{R099GF}) can be limited to the first order terms resulting in the straightforward relation:
\begin{equation}
	\label{R101GF}
	\left. \begin{array}{ll}
\displaystyle \chi^{(k)}_{i} = M_{ij}^{(k)} \xi^{(k)}_{j}.\\
\end{array} \right. 
\end{equation} 

The overall identification procedure include the following steps: 1) temperature dependence measurements of resonant frequencies and fitting of frequency-temperature plots to $3$rd order polynoms (experience shows that higher orders are not necessary). If possible applied this to three modes with wave numbers $(n,0,0)$, $(n,2,0)$ and $(n,0,2)$ with $n\geq3$; 2) application of correction (\ref{R004GF}) giving an eigenfrequency of an infinite plate; 3) calculation of temperature coefficients $\chi^{(k)}_{n}$ of this frequency; 4) recalculation of effective material temperature coefficients $\xi^{(1)}_{e}$ by solving (\ref{R101GF}). Note that the data set of $G^{(i)}_{j}$ values used in the following paragraphs is that identified at room temperature from \cite{MalochaIFCS2000}. According to (\ref{R002GF}), this approximation is largely justified because frequency changes never exceed a few $10^{-4}$ from room to cryogenic temperatures.


The temperature coefficients of the elastic tensor of cryogenically cooled LGT are identified using temperature measurements of high quality plate resonators. A set of resonators with different plate orientations with respect to different crystal orientation (crystal cuts) are needed to capture anisotropy of these parameters. The following crystal cuts in accordance with the IEEE-std 176-1987 \cite{angleIEEEstd} are used: Y ($90^\circ$, $0^\circ$), X ($0^\circ$, $0^\circ$), Z ($0^\circ$, $90^\circ$), X+45 ($0^\circ$, $45^\circ$), Y+15-30 ($105^\circ$, $-30^\circ$), Y-45 ($90^\circ$, $-45^\circ$). The corresponding Euler angles ($\phi$, $\theta$) are given in brackets. The full list of characterised resonators and corresponding modes is given in Table~\ref{modesT}. The measured modes belong to three classes: A$_{nmp}$ - thickness quasi-longitudinal, B$_{nmp}$ - thickness fast quasi-shear and C$_{nmp}$ - thickness slow quasi-shear.

The resonators are fabricated according to a process that is very close to the conventional process for quartz BAW devices: crystal orientation, crystal cut, plate rounding, lapping and polishing for a plano-convex surface type~\cite{Brice1985}. The resonators are designed to have a resonance frequency on the 3rd OT of the shear mode of about $10$~MHz at room temperature. The original crystal material is supplied by FOMOS-Materials (Russia)\cite{fomos}. 

\begin{table}[t]
\caption{Crystal plate cuts and characterised resonances. Temperature coefficients $\chi^{(k)}_{nmp}$ of resonance frequencies (\ref{R005GF}) of a set of resonators as a result of $3$rd order polynomial fit, with $T_{ref}=8K$.{ The mode subscript determines wave numbers as X$_{mnp}$. $A$ refers to the thickness-extensional mode. $B$ and $C$ are for both thickness shear modes.}
}
\centering
\begin{tabularx}{\columnwidth}{XXXXX}
\hline
\hline
Cut & X$_{nmp}$ & $\chi^{(1)}_{nmp},10^{-8}$  & $\chi^{(2)}_{nmp},10^{-8}$  & $\chi^{(3)}_{nmp},10^{-8}$ \\
\hline
 Y\footnotemark[1] & C$_{300}$& $+11.235$   & $+7.386$   &  $+1.164$ \\
   & C$_{320}$&  $-9.28$   & $+5.46$   & $+1.92$  \\
   & C$_{302}$&  $+10.21$   &  $+7.45$  &  $+1.08$ \\
 \hline
 Y\footnotemark[1] & C$_{500}$ & $+10.383$    & $+7.783$   &$+1.189$   \\
   & C$_{520}$&  $+7.28$   & $+7.20$   & $+1.135$  \\
   & C$_{502}$&   $+12.30$  & $+7.97$   & $+1.20$  \\
 \hline
 X\footnotemark[2] & A$_{300}$ & $-66.28$    &  $-7.57$  & $-0.150$  \\
  & A$_{500}$ & $-74.796$    & $-8.76$   &  $-0.091$  \\
\hline
 Z\footnotemark[3]& C$_{500}$& $+2.50$    & $+7.37$   & $+1.14$  \\
   & C$_{300}$& $+4.86$    & $+7.66$   & $+1.08$  \\
\hline
 X+45$^\circ$ &C$_{300}$ &  $+108.11$    & $+23.97$   & $+2.235$  \\
  & A$_{300}$ & $-55.69$     & $-6.419$   & $-0.1046$  \\
\hline
 Y+15$^\circ$-30$^\circ$ & B$_{500}$ & $-75.06$     & $-7.68$   & $0.1642$  \\
 & C$_{300}$ &  $+235.88$ & $+45.33$ & $3.63$ \\
\hline
 Y-45$^\circ$ & C$_{300}$ &  $+179.0$    & $+37.57$   & $+3.75$  \\
 & C$_{500}$ & $+454.8$     &  $+75.42$  & $+5.126$  \\
 & C$_{700}$ &  $+463.24$    & $+73.23$   & $+5.27$  \\
\hline
\end{tabularx}
\footnotetext[1]{available for infinite plate. See (\ref{R004GF}) }
\footnotetext[2]{shear waves could not be excited}
\footnotetext[3]{excited using an electrical lateral field instead of a normal field as for Y cuts}
\label{modesT}
\end{table}


The LGT resonators of different cuts have been cooled down to liquid helium temperatures using a conventional pulse tube cryocooler. The resonators without an enclosure, an individual vacuum chamber, are put inside an oxygen-free copper block whose temperature is controlled with $\pm3$mK precision. The resonators and the copper block are in thermal equilibrium at each measurement point. The resonators are connected to the measurement room temperature electronics using long coaxial cables thermalized at each stage of the cryocooler. Although resonators do not have their own vacuum chamber, the pressure of their environment is of the order of $5\cdot10^{-7}$~mbar. The environmental instabilities are negligible for the current measurements~\cite{Goryachev:2012aa}.

Several resonances (Table~\ref{modesT}) are characterised in the temperature range $3.75$-$15$K using the network analyser technique. The technique incorporates a careful calibration stage that allows the method to exclude influence of the connecting cables. For this purpose three calibration standards are installed in the very proximity to the characterised device. The technique employs a network analyser HP4195A locked to a Hydrogen Maser for the ultimate frequency stability. The instrumentation provides $0.001$dB, $0.01^\circ$ resolution in magnitude and phase respectively, and up to $1$mHz frequency precision.  
 Typically complex valued impedance $Z(f)$ is measured in the vicinity of different resonance frequencies at each temperature value (see Fig.~\ref{reson}). Typical frequency response of a LGT resonator in a vicinity of a resonance frequency is shown in Fig.~\ref{reson}. For each such measurement, parameters of the equivalent Butterworth-van Dyke model~\cite{Salt} are identified giving values of the resonance frequency, quality factor and active (motional) resistance. Then, a third order fit of various curves of resonance frequencies versus temperature provides estimated values of the frequency coefficients $\chi^{(k)}_{i}$, which are further used to identify the temperature coefficients of the elastic tensor.

\begin{figure}[t!]
	\centering
			\includegraphics[width=0.45\textwidth]{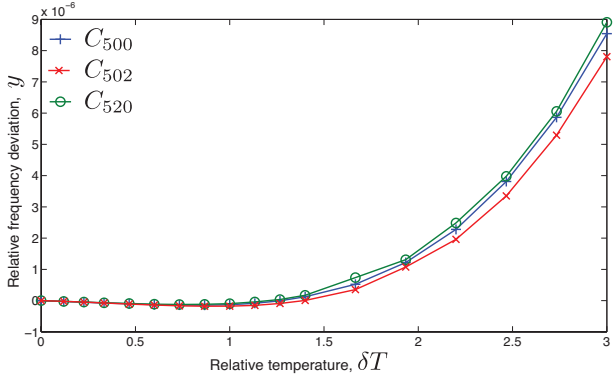}
	\caption{Dependence of the resonance fractional frequency shift $y = \frac{f(T)-f(T_\text{ref})}{f(T_\text{ref})}$ of an Y-cut LGT resonator as a function of temperature offset $\delta T = T-T_\text{ref}$.}
	\label{temperdep}
\end{figure}

Fig.~\ref{temperdep} demonstrates dependence of the Y-cut resonator fractional frequency shift against the temperature offset. The reference temperature is taken $3.75$K. The dependence shows a frequency-temperature turn over point corresponding approximately to $5$K. Table~\ref{coeffsT} gives the resulting temperature coefficients up to the third order of the elastic tensor elements $G_{ij}$. Coefficients $\xi^{(k)}_{14}$ are not given since they can be easily derived using the expression $\xi^{(k)}_{14} = \xi^{(k)}_{17}+\xi^{(k)}_{44}-\xi^{(k)}_{47}$. 

\begin{table}[t]
\caption{Temperature coefficients $\xi_{ij}^{(k)}$ of the elastic elements $G_{ij}$. In brackets: estimated uncertainties that are extracted from experimental results on a set of cuts. Beyond measurement uncertainty the dominant factor is the cut angle uncertainty ($\sim\pm3^\prime$).} 
\centering
\begin{tabularx}{\columnwidth}{XXXX}
\hline
\hline
Coefficient & $\xi_{ij}^{(1)}$, $10^{-6}$ &$\xi_{ij}^{(2)}$, $10^{-6}$ & $\xi_{ij}^{(3)}$, $10^{-7}$\\
\hline
$\xi_{11}^{(k)}$ & -1.4 $(20\%)$& -0.16 $(20\%)$& -0.024 $(20\%)$\\
$\xi_{66}^{(k)}$ & 0.273 $(415\%)$& 0.16 $(151\%)$& 0.22 $(116\%)$\\
$\xi_{55}^{(k)}$ & 0.074 $(20\%)$& 0.15 $(20\%)$& 0.22 $(20\%)$\\
$\xi_{17}^{(k)}$ & -21.33 $(20\%)$& -3 $(21\%)$& -1.73 $(24\%)$\\
$\xi_{13}^{(k)}$ & 0.4 $(33\%)$& 0.087 $(46\%)$& -0.09 $(163\%)$\\
$\xi_{33}^{(k)}$ & -29.8 $(23\%)$& -2.75 $(55\%)$& -0.73 $(157\%)$\\
$\xi_{44}^{(k)}$ & 86 $(23\%)$& 7.75 $(61\%)$& 2.08 $(159\%)$\\
$\xi_{47}^{(k)}$ & 33.3 $(24\%)$& 2.95 $(68\%)$& 0.73 $(153\%)$\\
 \hline
\end{tabularx}
\label{coeffsT}
\end{table}

{ Table~\ref{coeffsCOMP} compares temperature coefficients of the $G_{ij}$ matrix at liquid helium and room temperatures. It can be noted that while the first order coefficients at $5$K are significantly lower than at $298$K, the second order coefficients exhibit the inverse relation. The temperature sensitivities of mechanical parameters at higher temperatures, e.g. up to $900^\circ$C\cite{Sturtevant,Davulis2013}, are typically given for the the Euler formalism, and thus cannot be easily compared.}

\begin{table}[t]
\caption{Comparison of the temperature coefficients at the liquid helium temperature (this work) and $298$K\cite{Bourquin2006,Bourquin2009}.} 
\centering
\begin{tabularx}{\columnwidth}{XXXXX}
\hline
\hline
 & $\xi_{ij}^{(1)}$, $10^{-6}$ &$\xi_{ij}^{(1)}$, $10^{-6}$ & $\xi_{ij}^{(2)}$, $10^{-7}$& $\xi_{ij}^{(2)}$, $10^{-9}$\\
 & $4.2$K  & $298$K & $4.2$K & $298$K\\
\hline
$\xi_{11}^{(k)}$ & -1.4 & -68.9 & -0.16 & -74\\
$\xi_{66}^{(k)}$ & 0.273 & 20.7 & 0.16  & -172\\
$\xi_{55}^{(k)}$ & 0.074 & 13.0 & 0.15  & -143\\
$\xi_{17}^{(k)}$ & -21.33 & -391 & -3  & 285 \\
$\xi_{13}^{(k)}$ & 0.4 & -76.6 & 0.087  & -92\\
$\xi_{33}^{(k)}$ & -29.8 & -109 & -2.75 & -72.7\\
$\xi_{44}^{(k)}$ & 86 & 8.4 & 7.75  & -147\\
$\xi_{47}^{(k)}$ & 33.3 & 10.7 & 2.95 & -145\\
 \hline
\end{tabularx}
\label{coeffsCOMP}
\end{table}

In order to ensure linear response of the device the resonators are characterised at the lowest accessible incident power, $-50$dBm. The typical power sensitivity of a resonance frequency of an LGT resonator is shown in Fig.~\ref{powersens}. The resonators demonstrate expected Duffing-type nonlinearity accompanied by thermal effects at elevated values of incident power with no anomalous nonlinear behaviour\cite{quartzJAP}. 

\begin{figure}[t!]
	\centering
			\includegraphics[width=0.45\textwidth]{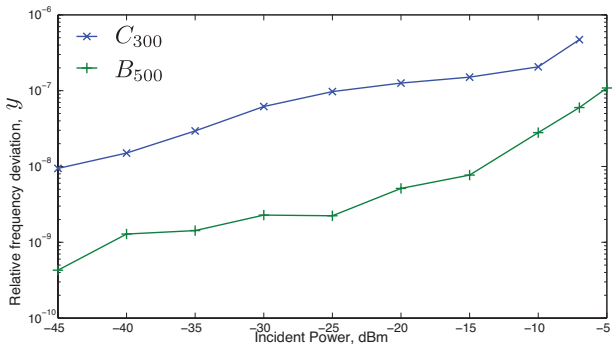}
	\caption{Typical power sensitivity of the resonance frequency $y = \frac{f(P)-f(-50\text{dBm})}{f(-50\text{dBm})}$ of an LGT resonator  at the liquid helium temperature.  The curves are given for shear modes of the Y$+15^\circ-30^\circ$ cut LGT resonator. }
	\label{powersens}
\end{figure}


Resonator characterisation at different temperatures gives not only frequency change but also a temperature dependence of acoustical losses or equivalently quality factors of acoustic resonances. To determine such dependence, two Y-cut resonators have been measured in a wide range of cryogenic temperatures. Unlike their quartz counterparts\cite{Goryachev1,ScRep,quartzPRL}, none of the tested LGT resonators exhibit significant improvement of their quality factor. Its slight improvement (less than an order of magnitude from about $10^6$ to $10^{7}$) demonstrates no power law in temperature dependence. These facts show that device losses are most likely dominated by engineering losses, i.e. clamping problem, due to lack of phonon trapping. Thus, regimes of material losses for LGT resonators have not been achieved yet. This is confirmed by the considerable difference in quality factors for two resonators resulting from non-reproducibility of the technology. Additionally, the two resonators demonstrate a sharp peak of acoustic wave absorption near $40$K. Similar to quartz resonators, this could be related to impurity ions inevitably found in the crystalline structure~\cite{gfact}, {or more probably due to activity dips\cite{Johnson2002}, because peaks of both samples do not appear at the same temperature. }

\begin{figure}[t!]
	\centering
			\includegraphics[width=0.45\textwidth]{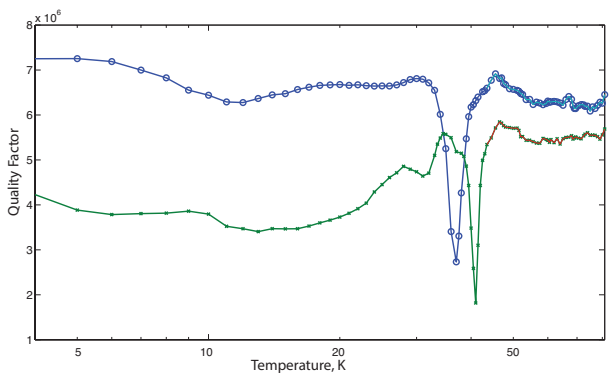}
	\caption{Temperature dependence of quality factors of two Y-cut resonators of the same type in a wide range of temperatures. }
	\label{powersens}
\end{figure}


In summary, various cuts of LGT crystal resonators have been tested at liquid helium temperatures. Thermal coefficients of the elastic tensor in Lagrangian representation are calculated. 
Measurements revealed existence of the frequency-temperature points for slow shear modes of Z and Y cuts at around $7$K. Possibility of such points is vital for building temperature insensitive devices for ultra-stable frequency reference systems. It may lead to improve stability of cryogenic BAW frequency sources\cite{perspect} Power sensitivity of LGT-based devices demonstrates typical Duffing-type nonlinearity and thermal effects\cite{SunFr,quartzJAP}. Loss regimes of the devices under study are determined by clamping losses due to lack of effective phonon trapping~\cite{Galliou:2008ve,Goryachev1,ScRep}. Nevertheless, since the piezoelectric coupling coefficient well exceeding that of quartz, low loss LGT is a promising material for various physical applications at low temperatures including frequency control, hybrid quantum systems, sensing applications.

{The obtained results can be used further to investigate thermal sensitivity of the LGT based devices. On one hand, calculation of cuts with suitable frequency-temperature turnover points may lead to improve stability of cryogenic BAW frequency sources\cite{perspect}. On another hand, the same data can be used to increase temperature sensitivity for sensor applications. So, with piezoelectric coupling coefficient well exceeding that of quartz, low loss LGT is a promising material for various physical applications at low temperatures including frequency control, hybrid quantum systems, sensing applications.}


This work is supported by Conseil R\'egional de Franche-Comt\'e (Convention No. 2008C 16215). MG is thankful to the Australian Research Council under grant CE110001013.

\section*{References}

%


\end{document}